\input harvmac.tex

\lref\pleb{J.F. Pleba\'nski, ``On the separation of Einstenian
substructures'', {\it J. Math.  Phys.} {\bf 18}, 2511 (1977).}

\lref\pleban{J.F. Pleba\'nski, ``Some solutions of complex Einstein
equations'', {\it J. Math. Phys.} {\bf 16}, 2395 (1975).}

\lref\capo{R. Capovilla, J. Dell, T. Jacobson and L. Mason, {\it Class. Quant.
Grav.} {\bf 8}, 41 (1991).}

\lref\ashtekar{A. Ashtekar and R.S. Tate, {\it Lectures on
Non-perturbative Canonical Gravity}, (World Scientific, Singapore, 1991).}

\lref\hh{C.P. Boyer, J.D. Finley III, and J.F. Pleba\'nski, ``Complex
General Relativity, ${\cal H}$ and ${\cal HH}$ Spaces- A Survey of One
Approach'', in  A.
Held, ed., {\it General Relativity and Gravitation}, Vol. 2, (Plenum Press, New York, 
1980), pp.241-281.}

\lref\penrose{R. Penrose, ``Nonlinear Gravitons and Curved Twistor
Theory'', {\it Gen. Rel.  Grav.} {\bf 7}, 31 (1976).}

\lref\newman{M. Ko, M. Ludvigsen, E.T. Newman and K.P. Tod, ``The Theory
of ${\cal H}$-Space'', {\it Phys. Rep.} {\bf 71}, 51 (1981).}

\lref\oova{H. Ooguri and C. Vafa, ``Self-duality and ${\cal N}=2$ String
Magic'', {\it Mod. Phys.  Lett.} A {\bf 5}, 1389 (1990); ``Geometry of ${\cal
N}=2$ Strings'', {\it Nucl. Phys.} {\bf 361}, 469 (1991).}

\lref\ovhet{ H. Ooguri and C. Vafa, ``${\cal N}=2$ Heterotic Strings'',
{\it Nucl. Phys.} {\bf 367}, 83 (1991).}

\lref\nunes{ A. Jevicki, M. Mihailescu, J.P. Nunes, ``Large $N$ WZW Field Theory of
${\cal N}=2$ Strings'', {\it Phys. Lett.} B {\bf 416}, 334 (1998), hep-th/9706223;
``Large $N$ Field Theory of ${\cal N}=2$ Strings and Self-dual Gravity'', {\it Chaos
Solitons Fractals} {\bf 10}, 385 (1999), hep-th/9804206.}

\lref\ccastroone{ C. Castro, ``${\cal N}=2$ Super-Wess-Zumino-Novikov-Witten Model Valued
in Superdiffeomorphisms SDiff$(M^2)$ is Self-dual Supergravity in Four Dimensions'', 
{\it J. Math. Phys.} {\bf 35}, 920 (1994).}

\lref\ccastrotwo{ C. Castro, ``${\cal W}$ Gravity, ${\cal N}=2$ Strings and $2+2$
SU$(\infty)$ Yang-Mills Instantons'', {\it J. Math. Phys.} {\bf 35}, 920 (1994).}

\lref\ade{M. Ademollo, L. Brink, A. D'Adda, R. D'Auria, E. Napolitano, S. Sciuto, 
E. Del Giudice, P. Di Vecchia, S. Ferrara, F. Gliozzi, R. Musto, R. Pettorino and J.H.
Schwarz, {\it Phys. Lett.} B {\bf 62}, 105 (1976); {\it Nucl.
Phys.} B {\bf 111}, 77 (1976).}

\lref\marcus{N. Marcus, ``A Tour Through ${\cal N}=2$ Strings'', in 
M. Bianchi, F. Fucito, E. Marinari and Augusto Sagnotti, eds.,
{\it String Theory, Quantum Gravity and the Unification of the Fundamental
Interactions}, (World Scientific, Singapore, 1993), hep-th/9211059.}

\lref\pam{ O. Lachtenfeld, ``Mathematics and Physics of ${\cal N}=2$
Strings'', hep-th/9912281.}

\lref\gatesone{ S. James Gates Jr., ``Superspace Formulation of New Nonlinear Sigma
Models'', {\it Nucl. Phys.} B {\bf 238}, 349 (1984).}

\lref\gatestwo{ S. James Gates Jr., C.M. Hull and M. Rocek, ``Twisted Multiplets and New
Supersymmetric Nonlinear Sigma Models'', {\it Nucl. Phys.} B {\bf 248}, 157 (1984).}

\lref\hull{C.M. Hull, ``The Geometry of ${\cal N}=2$ Strings with
Torsion'', {\it Phys. Lett.} B {\bf 387}, 497 (1996), hep-th/9606190.}

\lref\bv{N. Berkovits and C. Vafa, ``${\cal N}=4$ Topological Strings'',
{\it Nucl. Phys.} {\bf 433}, 123 (1995).}

\lref\ovdos{H. Ooguri and C. Vafa, ``All Loop ${\cal N}=2$ String
Amplitudes'', {\it Nucl. Phys.} {\bf 451}, 121 (1995), hep-th/9505183.}

\lref\npb{ N. Marcus, ``The ${\cal N}=2$ Open String'', {\it Nucl. Phys.} 
{\bf 387}, 263 (1992), hep-th/9207024.}

\lref\ketov{ S.V. Ketov, ``Mixed (Open/Closed) ${\cal N}=(2,2)$ String
Theory as an Integrable Deformation of Self-dual Gravity'',
hep-th/9612170; ``From ${\cal N}=2$ Strings to F and M Theory'', hep-th/9606142.}

\lref\ahs{M.F. Atiyah, N.J. Hitchin and I.M. Singer, ``Self-duality in
Four-dimensional Riemannian Geometry'', {\it Proc. Roy. Soc. Lond.} A {\bf 362},
425 (1978).}

\lref\yang{ C.N. Yang, {\it Phys. Rev. Lett.} {\bf 38}, 1377 (1977).}

\lref\new{ J.F. Pleba\'nski and M. Przanowski, ``A New Version of the
Heavenly Equation'', in Alex Harvey, ed., {\it On Einstein's Path: Essays in Honor of
Engelbert Schucking}, (Springer Verlag, Berlin, 1999), pp.
381-405.}

\lref\aefectiva{D. Gl\"uck, Y. Oz, and T. Sakai, ``The Effective Action
and Geometry of Closed ${\cal N}=2$ Strings'', {\it JHEP} {\bf 0307}, 0007(2003),
hep-th/0304103.}

\lref\giveon{ A. Giveon and M. Rocek, ``On the BRST Operator Structure of
${\cal N}=2$ String'', hep-th/9302049.}

\lref\stringh{ O. Lechtenfeld and A.D. Popov, ``Closed ${\cal N}=2$
Strings: Picture Changing, Hidden Symmetries and SDG Hierarchy'', {\it Int. J.
Mod. Phys.} A {\bf 15}, 4191 (2000), hep-th/9912154.}

\lref\bischoff{ J. Bischoff, S.V. Ketov and O. Lechtenfeld, {\it Nucl. Phys.} B 
{\bf 438}, 373 (1995), hep-th/9406101.}

\lref\nsfketov{ S.V. Ketov and O. Lechtenfeld, {\it Phys. Lett.} B {\bf 353}, 463
(1995), hep-th/9503232.}

\lref\gr{E. Witten, ``Ground Ring of Two-dimensional String Theory'',
{\it Nucl. Phys.} {\bf B373}, 187 (1992).}

\lref\wz{ E. Witten and B. Zwiebach, ``Algebraic Structures and
Differential Geometry in Two-dimensional String Theory'', {\it Nucl. Phys.} B
{\bf 377}, 55 (1992).}

\lref\lz{B.H. Lian and G.J. Zuckerman, ``New Perspectives on the BRST
Algebraic Structure of String Theory'', {\it Commun. Math. Phys.} {\bf 154},
613 (1993), hep-th/9211072.}

\lref\groundring{ H. Garcia-Compe\'an, M. Przanowski and J.F Pleba\'nski,
``On Self-dual Gravity Structures on Ground Ring Manifolds in
Two-dimensional String Theory'', {\it Rev. Mex. Fis.}  {\bf 40}, 547
(1994), hep-th/9405081.}

\lref\bpone{C.P. Boyer and J.F. Pleba\'nski, {\it J. Math. Phys.} {\bf 18},
1022 (1977).}

\lref\bptwo{ C.P. Boyer and J.F. Pleba\'nski, ``An Infinite Hierarchy of
Conservation Laws and Nonlinear Superposition Principles for Self-dual
Einstein Spaces'', {\it J. Math. Phys.} {\bf 26}, 229 (1985).}

\lref\mirror{ E. Witten, ``Mirror Manifolds and Topological Field
Theory'', hep-th/9112056.}

\lref\aspin{P.S. Aspinwall, ``K3 Surfaces and String Duality'',
hep-th/9611137; P.S.  Aspinwall and D.R. Morrison, ``String Theory on K3
Surfaces'', hep-th/9404151.}

\lref\inst{ O. Lechtenfeld, ``${\cal N}=2$ Worldsheet Instantons Yield
Cubic Self-dual Yang-Mills'', {\it Phys. Lett.} B, {\bf 405}, 49 (1997).}

\lref\lezparkes{A.N. Leznov, Teor. M. Phys. {\bf 73}, 1233 (1988); A.
Parkes, {\it Phys. Lett.} B {\bf 286}, 265 (1992).}

\lref\parkes{A. Parkes, ``On ${\cal N}=2$ Strings and Classical Scattering
Solutions of SD Yang-Mills in (2,2) Space'', {\it Nucl. Phys.} B {\bf 376},
279 (1992).}

\lref\cswitt{E. Witten, ``Chern-Simons Gauge Theory as a String Theory'',
{\it Prog. Math.} {\bf 133}, 637 (1995), hep-th/9207094.}

\lref\gopa{R. Gopakumar and C. Vafa, ``Topological Gravity as a Large-$N$
Topological Gauge Theory'', {\it Adv. Theor. Math. Phys.} {\bf 2}, 413 (1998), 
hep-th/9802016.}

\lref\mmarino{M. Mari\~no, ``Enumerative Geometry and Knot Invariants'',
hep-th/0210145.}

\lref\kg{M. Bershadsky and V. Sadov, ``Theory of Kahler Gravity'', {\it Int. J.
Mod. Phys.} A {\bf 11}, 4689 (1996), hep-th/9410011.}

\lref\chiral{J.F. Pleba\'nski, M. Przanowski and H. Garcia-Compe\'an,
``From Principal Chiral Model to Self-dual Gravity'', {\it Mod. Phys. Lett.} A
{\bf 11}, 663 (1996), hep-th/9509092.}

\lref\nappi{C. Nappi, ``Some Properties of an Analog of the Chiral
Model'', {\it Phys. Rev.} D {\bf 21}, 418 (1980).}

\lref\plebprzan{J.F. Pleba\'nski and M. Przanowski, ``The Lagrangian of a
Self-dual Gravitational Field as a Limit of the SDYM Lagrangian'', {\it Phys.
Lett.} A {\bf 212}, 22 (1996), hep-th/9605233.}

\lref\martinec{ E. Martinec, ``M-Theory and ${\cal N}=2$ Strings'',
hep-th/9710122.}

\lref\gatesthree{ S. James Gates Jr., L. Lu and R.N. Oerter, ``Simplified SU(2) 
Spinning String Superspace Supergravity'', {\it Phys. Lett.} B {\bf 218}, 33 (1989).}

\lref\cheung{ Y.K.E. Cheung, Y. Oz and Z. Yin, ``Families of ${\cal N}=2$
Strings'', {\it JHEP} {\bf 0311}, 026 (2003), hep-th/0211147.}

\lref\dbranes{ D. Gl\"uck, Y. Oz, and T. Sakai, ``D-branes in ${\cal N}=2$
Strings'', {\it JHEP} {\bf 0308}, 055 (2003), hep-th/0306112.}

\lref\hhone{J.F. Pleba\'nski and I. Robinson, ``Left-Degenerate Vacuum
Metrics'', {\it Phys.  Rev. Lett.} {\bf 37}, 493 (1976).}

\lref\toppan{Z. Popowicz and F. Toppan, ``The ${\cal N}=2$ Supersymmetric Heavenly
Equation and Its Super-Hydrodynamical Reduction'', {\it J. Phys.} A {\bf 36},
9701 (2003), nlin.si/0302061.}

\lref\twistor{E. Witten, ``Perturbative Gauge Theory as a String Theory in
Twistor Space'', {\it Commun. Math. Phys.} {\bf 252}, 189 (2004), hep-th/0312171.}

\lref\berko{N. Berkovits, ``Self-dual SuperYang-Mills as a String Theory
in $(x,\theta)$ Space'',  {\it JHEP} {\bf 0405}, 034 (2004), hep-th/0403280.}

\let\includefigures=\iftrue
\newfam\black
\includefigures
\input epsf
\def\figin{\epsfcheck\figin}\def\figins{\epsfcheck\figins}
\def\epsfcheck{\ifx\epsfbox\UnDeFiNeD
\message{(NO epsf.tex, FIGURES WILL BE IGNORED)}
\gdef\figin##1{\vskip2in}\gdef\figins##1{\hskip.5in}
\else\message{(FIGURES WILL BE INCLUDED)}%
\gdef\figin##1{##1}\gdef\figins##1{##1}\fi}
\def\DefWarn#1{}
\def\figinsert{\goodbreak\midinsert}
\def\ifig#1#2#3{\DefWarn#1\xdef#1{fig.~\the\figno}
\writedef{#1\leftbracket fig.\noexpand~\the\figno}%
\figinsert\figin{\centerline{#3}}\medskip\centerline{\vbox{\baselineskip12pt
\advance\hsize by -1truein\noindent\footnotefont{\bf Fig.~\the\figno:} #2}}
\bigskip\endinsert\global\advance\figno by1}
\else
\def\ifig#1#2#3{\xdef#1{fig.~\the\figno}
\writedef{#1\leftbracket fig.\noexpand~\the\figno}%
\global\advance\figno by1}
\fi
\font\cmss=cmss10 \font\cmsss=cmss10 at 7pt

\def\IB{\relax\hbox{$\inbar\kern-.3em{\rm B}$}}
\def\IC{\relax\hbox{$\inbar\kern-.3em{\rm C}$}}
\def\IQ{\relax\hbox{$\inbar\kern-.3em{\rm Q}$}}
\def\ID{\relax\hbox{$\inbar\kern-.3em{\rm D}$}}
\def\IE{\relax\hbox{$\inbar\kern-.3em{\rm E}$}}
\def\IF{\relax\hbox{$\inbar\kern-.3em{\rm F}$}}
\def\IG{\relax\hbox{$\inbar\kern-.3em{\rm G}$}}
\def\IGa{\relax\hbox{${\rm I}\kern-.18em\Gamma$}}
\def\IH{\relax{\rm I\kern-.18em H}}
\def\IK{\relax{\rm I\kern-.18em K}}
\def\IL{\relax{\rm I\kern-.18em L}}
\def\IP{\relax{\rm I\kern-.18em P}}
\def\IR{\relax{\rm I\kern-.18em R}}
\def\Z{\relax\ifmmode\mathchoice
{\hbox{\cmss Z\kern-.4em Z}}{\hbox{\cmss Z\kern-.4em Z}}
{\lower.9pt\hbox{\cmsss Z\kern-.4em Z}}
{\lower1.2pt\hbox{\cmsss Z\kern-.4em Z}}\else{\cmss Z\kern-.4em
Z}\fi}
\def\IZ{Z\!\!\!Z}
\def\II{\relax{\rm I\kern-.18em I}}




\def\bar{\overline}




\def\inbar{\,\vrule height1.5ex width.4pt depth0pt}


\def\bar{\overline}

\def\example#1{\bgroup\narrower\footnotefont\baselineskip\footskip\bigbreak
\hrule\medskip\nobreak\noindent {\bf Example}. {\it #1\/}\par\nobreak}
\def\endexample{\medskip\nobreak\hrule\bigbreak\egroup}


\Title{\vbox{\baselineskip12pt
\hbox{CINVESTAV-FIS-22/04}
\hbox{hep-th/yymmddd}
}} {\vbox{
\centerline{${\cal N}$=2 String Geometry and the Heavenly Equations}}}
\centerline{Hugo Garcia-Compe\'an\foot{{\it ICTP Associated Member, The Abdus Salam 
International Centre for Theoretical Physics, Trieste, Italy}, E-mail Address:  
compean@fis.cinvestav.mx}}
\medskip
\centerline{\it Departamento de Fisica}
\centerline{\it Centro de Investigaci\'on y de Estudios Avanzados del IPN}
\centerline{\it Apdo. Postal 14-740, 07000, M\'exico D.F., M\'exico}

\vskip 10pt

\centerline{\it This paper is dedicated to Professor Jerzy F. Pleba\'nski on the occasion of his 75th 
birthday}

\vskip 20 pt

{\bf \centerline{Abstract}}
\noindent
In this paper we survey some of the relations between Pleba\'nski description of
self-dual gravity through the {\it heavenly equations} and the physics (and
mathematics) of ${\cal N}=2$ Strings. In particular we focus on the
correspondence between the infinite hierarchy in the ground ring structure of
BRST operators and its associated Boyer-Pleba\'nski construction of infinite
conserved quantities in self-dual gravity. We comment on ``Mirror Symmetry'' in
these models and the large-$N$ duality between topological ${\cal N}=4$ gauge
theories in two dimensions and topological gravity in four dimensions. Finally
D-branes in this context are briefly outlined.

\smallskip

\Date{May, 2004}


\newsec{Introduction}

The description of the gravitational field at the quantum level is nowadays one
of the biggest problems in theoretical physics. At the present time, there are
two contending approaches to quantum gravity: {\it Loop Quantum Gravity} and {\it
String Theory}. In the former one, the basic degrees of freedom of the quantum
gravitational field are represented by tinny loops, while in the latter one,
they are the fundamental strings or D-branes. These approaches are quite
different and a relation between them does not exist at the moment\foot{
Actually both approaches are conceptually different. In the string theory
approach, perturbative nonrenormalizability of general relativity suggest that
it is an effective field theory and it doesn't represent a fundamental theory.
Thus, new physical degrees of freedom can be found at higher energies in an
extended theory whose low energy limit gives precisely Einstein theory or some
generalizations of it. Loop Quantum Gravity suggest that the problem can be
faced in a strongest way by finding, at the nonperturbative level, a suitable
quantum theory for a Lagrangian which couples general relativity and matter.}.
In the present paper we focus on the latter one.

On the other hand, two of the great insights made by Pleba\'nski were two papers
published in the middle of seventies \refs{\pleb,\pleban}. In the first paper he
proposed a SL$(2,\IC)$ chiral Lagrangian for the classical degrees of freedom of
self-dual gravity. Nowadays, it is well known that this Lagrangian is related
through a Legendre transformation to a Hamiltonian approach \refs{\capo} which
is written in terms of the Ashtekar's variables \ashtekar.  The second paper
\pleban\ is perhaps less known among the loop quantum gravity people, however it
is very well know in the community of mathematical physicists working in exact
solutions of Einstein equations, in the theory of integrable systems and in
twistor theory. In this paper, Pleba\'nski found a description of self-dual
gravity degrees of freedom in terms of the so called {\it first heavenly
equation} in the ``weak heavens'' gauge or ${\cal B}$-gauge

\eqn\fhe{ \Omega_{,pr}({\bf x}) \Omega_{,qs}({\bf x}) - \Omega_{,ps}({\bf x})  
\Omega_{,qr}({\bf x}) = 1,}
where ${\bf x} \equiv(p,q,r,s)$ and $\Omega_{,pr} \equiv {\partial
\Omega({\bf x}) \over \partial p \partial r}$ etc. This is a second order
nonlinear differential equation satisfied by a holomorphic function $\Omega =
\Omega (p,q,r,s)$ in a particular coordinate system $\{ {\bf x} \} =
\{p,q,r,s\}$ of a complex four-dimensional manifold $M$. This equation
expresses the condition of Ricci-flatness of the {\it self-dual manifold} $M$
(or ${\cal H}$-space) and it is an equation for the K\"ahler potential
$\Omega({\bf x})$. A solution of this equation determines a self-dual metric
on $M$ which coincides precisely with the K\"ahler metric obteined from the
K\"ahler potential.

In the same reference \pleban\ it is given another description of self-dual
gravity in a different gauge. This is given by the Pleba\'nski {\it second}
heavenly equation

\eqn\she{
\Theta_{,xx}(\widetilde{\bf x}) \Theta_{,yy}(\widetilde{\bf x}) -
(\Theta_{,xy})^2(\widetilde{\bf x}) + \Theta_{,xp}(\widetilde{\bf
x}) + \Theta_{,yq}(\widetilde{\bf x}) = 0,}
where $\widetilde{\bf x} \equiv(x,y,p,q).$ This is also a nonlinear second order partial
differential equations for a holomorphic function $\Theta=\Theta(x,y,p,q)$ with $\{
\widetilde{\bf x} \} = \{p,q,x,y\}$ being another coordinate system in the ${\cal A}$-gauge,
where $x \equiv \Omega_{,p}$ and $y \equiv \Omega_{,q}$.

Both are descriptions of the ${\cal H}$-space in different {\it ``gauges''},
thus classically both descriptions are equivalent but they might give rise to
two inequivalent quantum descriptions. Various results about ${\cal H}$ and some
generalizations of them involving null strings called {\it Hyperheavenly} spaces
or ${\cal HH}$-spaces for short, are collected in the very nice survey paper
\hh.  Such a description has been proved to be equivalent to the Penrose
nonlinear graviton \penrose\ and that of the Newman approach reviewed in Ref.  
\newman.

In 1990 in a very important paper, Ooguri and Vafa \oova\ found that the first heavenly
equation was in fact, encoded in the description of string theory with local ${\cal
N}=2$ supersymmetry in the worldsheet. Soon after, in Ref. \ovhet, they were able to
the find a string theory (called the ${\cal N}=2$ heterotic string theory) such that
the target space effective field theory consists of self-dual gravity coupled to
self-dual Yang-Mills theory, just as in the ordinary heterotic string theory in ten
dimensions. A different description of ${\cal N}=2$ strings in terms of a large $N$
limit of a WZW field theory was proposed in Ref. \nunes. The connection
between ${\cal N}=2$ self-dual supergravity and ${\cal N}=2$ super-WZW was
worked out in Ref. \ccastroone. ${\cal N}=2$ strings have been also studied in relation to 
${\cal W}$ gravity and SU$(\infty)$ Yang-Mills instantons \ccastrotwo.

The present paper is dedicated to Professor Jerzy F. Pleba\'nski on the occasion
of his 75th birthday. Although exact solutions of the Einstein equations have
been his constant preoccupation in mathematical-physics, he has also made some
of the leading contributions to self-dual gravity whose formalism was shown to
be relevant to describe the geometry of ${\cal N}=2$ strings. I hope that this
modest
contribution to understand these systems presented here seems to be appropriate for
this occasion.

\subsec{${\cal N}=2$ Strings and Scattering Amplitudes}

\noindent
{\it Set up}

The purpose of this section will be to survey some geometry of the complex
spacetime Einstein self-dual geometry (${\cal H}$-spaces) viewed as the target
space geometry in the context of the worldsheet theory of ${\cal N}=2$ strings.
This theory is, in fact, a ${\cal N}=2$ supersymmetric nonlinear-sigma model
with {\it flat} target space and it was proposed by the first time in Ref. \ade.  
We will follows mainly in Ref. \oova, however some very nice expository articles
can be found in Refs. \refs{\marcus,\pam}. We will focus mainly in closed
strings, however some comments for open strings will be given.

Closed string theory with ${\cal N}=2$ worldsheet local supersymmetry is given
by ${\cal N}=2$ supergravity in two dimensions coupled to ${\cal N}=2$
superconformal matter. This theory is based on an action which has ${\cal N}=2$
superconformal symmetry. For closed strings there are two infinite algebras
acting independently on left- and right-sectors whose generators are
$(G,G^*,T,J)_L$ and $(\overline{G},\overline{G}^*,\overline{T},\overline{J})_R$,
respectively. The quantization of ${\cal N}=2$ string theory with these infinite
symmetries and the appropriate counting of ghost fields leads to an anomaly free
theory (total central charge vanishing) whose critical dimension $D = 2_{\IC}$.
That means that target space $M^D$ has two complex dimensions and target space
can have signature $(2,2)$ or $(0,4)$. In the present paper we will focus in
backgrounds with signature $(2,2)$, however, occasionally we make some comments
on the theory with $(0,4)$ signature.  The spectrum for this theory consists of
only one complex scalar field on the target space $M^{2,2}$. All oscillator
excitations are vanishing and there is not spacetime supersymmetry.

After gauge fixing all supergravity fields we keep with a theory for a ${\cal N}=2$
superfield $X: {\cal S} \to M^D$ which is the ${\cal N}=2$ supermatter and ${\cal S}$
is a ${\cal N}=2$ super-Riemann surface. This theory is based on the action
\eqn\action{ S_0 = \int_{\cal S} {d^2 z\over \pi} d^2 \theta d^2 \overline{\theta}
K_0(X, \overline{X}), } 
where $X^i$ ($i=1,2$) are two chiral superfields given by
$X^i(z, \bar{z}; \theta, \bar{\theta}) = x^i(z,\bar{z}) + \psi^i_R(z, \bar{z}) \theta^-
+ \psi^i_L(z, \bar{z}) \bar{\theta}^- + F^i(z,\bar{z})\theta^- \bar{\theta}^-$, where
$z = x - \theta^+ \theta^-$ and $K_0(X, \overline{X}) = X^1 \bar{X}^1 - X^2 \bar{X}^2$
is the K\"ahler potential.

For ${\cal N}=2$ matter we have two possibilities: $(i)$ Two chiral
superfields $X^i$ (and $\bar{X}^{\bar{i}}$) ($i=1,2$ and
$\bar{i}=\bar{1},\bar{2}$) where the fermions $\psi^i_L(z, \bar{z})$ and
$\psi^i_R(z, \bar{z})$ are charged under the U(1) group. The chiral
superfields satisfy $\overline{D}_{\pm}X^i = 0,$ and ${D}_{\pm}
\bar{X}^{\bar{i}} = 0$ and $(ii)$ two chiral superfields $X^i$ ($i=1,2$)
and two {\it twisted} chiral superfields\foot{ The existence of twisted chiral superfields
was pointed out by the first time in Ref. \gatesone . The name of ``twisted
chiral multiplets''
was first introduced in Ref. \gatestwo.} $\Sigma^a$ ($a=1,2$), satisfying:
$\overline{D}_{\pm}X^i = 0,$ ${D}_{\pm} \bar{X}^{\bar{i}} = 0$ and
${D}_{+}\Sigma^a = 0,$ $\overline{D}_{-}\Sigma^a = 0,$ ${D}_{-}
\overline{\Sigma}^{\overline{a}} = 0,$ and $\overline{D}_{+}
\overline{\Sigma}^{\overline{a}} = 0.$ The first selection leads to the
description of the usual self-dual geometry \oova. The second possibility leads to
the description of a self-dual geometry with hermitian
(non-Kahlerian) structure and torsion \refs{\gatestwo,\hull}.

\vskip .5truecm
\noindent
{\it Scattering Amplitudes of Closed, Open and Open/Closed ${\cal N}=2$
Strings}

For the closed string the only non-vanishing scattering amplitudes is the three-point
function on the sphere (fig. 1) \oova. This is given by:

\eqn\threep{
{\cal A}_{ccc} = \bigg\langle V_c|_{\theta =0}(0) \cdot \int_{\cal S} d^2 \theta d^2 \bar{\theta} 
V_c(1) 
\cdot 
V_c|_{\theta=0}(\infty) \bigg\rangle = \kappa c^2_{12} \not= 0,
}
where $c_{12} \equiv k_1 \bar{k}_2 - \bar{k}_1 k_2$ and  $V_c= {\kappa 
\over \pi} \exp \big( k \cdot \bar{X} + \bar{k} \cdot X\big)$ is the
vertex operator for the complex scalar field $X$.

\ifig\sphere{Three-point scattering amplitude on the sphere is the only non-vanishing
amplitude.}
{\epsfxsize2.2in\epsfbox{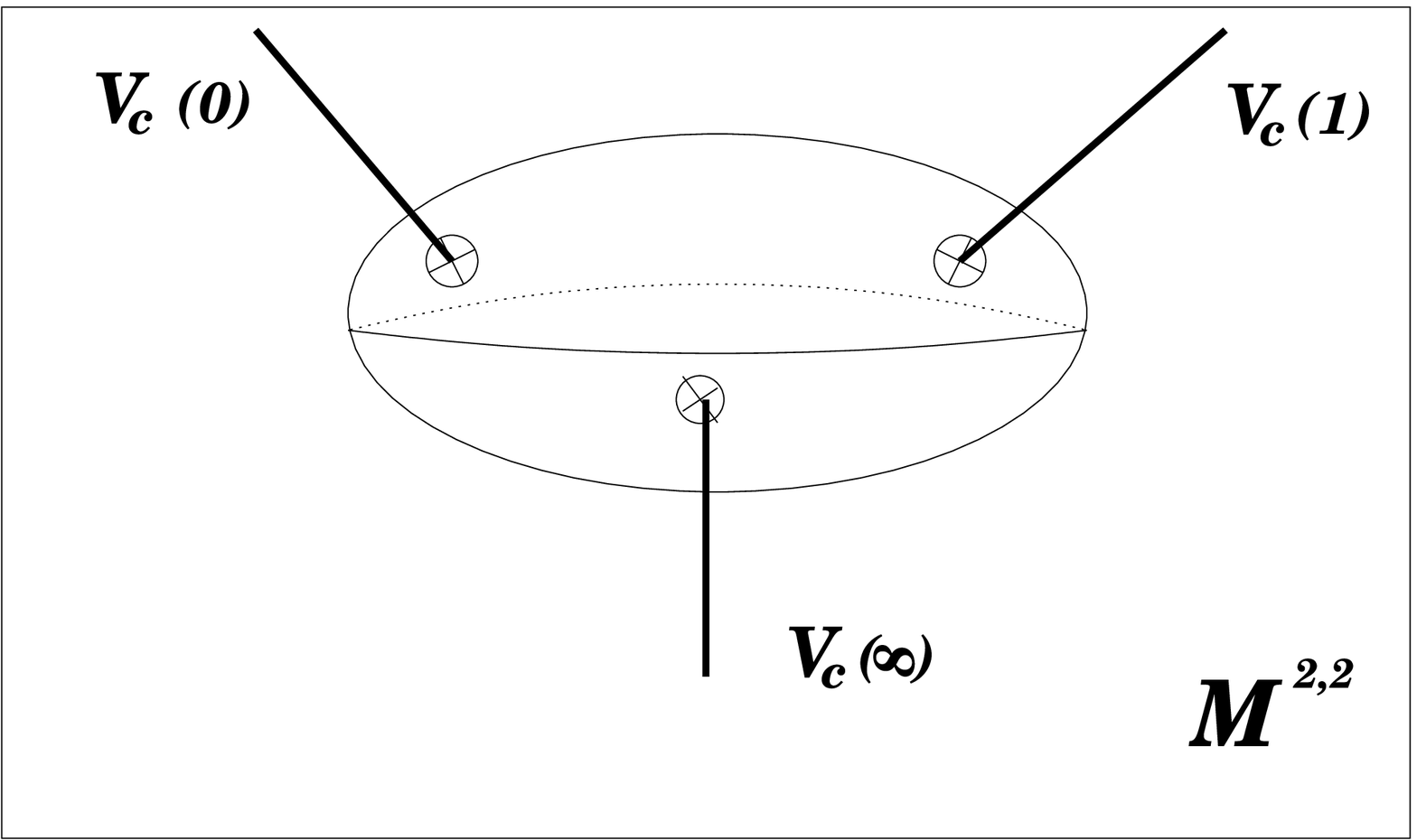}}


Four point function is given by

\eqn\fourp{
{\cal A}_{cccc} = \bigg\langle V_c|_{\theta =0}(0) \cdot \int_{\cal S} d^2 \theta d^2 \bar{\theta} 
V_c(z) 
\cdot \int_{\cal S} d^2 \theta d^2 \bar{\theta} V_c(1) \cdot
V_c|_{\theta=0}(\infty) \bigg\rangle,
}
which is vanishing on-shell. Higher order amplitudes for higher point and
higher genus 
correlation functions are vanishing as well \refs{\bv,\ovdos}.

For open strings the story is quite similar. Three-point scattering
amplitude at the tree level (on the disk) is given by \npb\

\eqn\threeo{
{\cal A}_{ooo} = \bigg\langle V_o|_{\theta =0}(0) \cdot \int d^2 \theta  V_o(1) \cdot
V_o|_{\theta=0}(\infty) \bigg\rangle = g c_{12} (-i f^{abc}),
}
where $a,b,c$ stand for the Chan-Paton indices and $V_o \equiv g \exp(k
\cdot \bar{X} + \bar{k} \cdot X)$ is the corresponding vertex operator.
Four-point scattering amplitude ${\cal A}_{oooo}$ vanishes on-shell as
well for the same reasons.

For the case of mixing open/closed scattering amplitudes we have for the
case of three-point function \ketov\

\eqn\threem{
{\cal A}_{ooc}  \propto \int_{- \infty}^{+ \infty} dx \bigg\langle V_o|_{\theta =0}(x) \cdot \int_{\cal 
S} d^2 \theta d^2 \bar{\theta} V_c(z=i) \cdot
V_o|_{\theta=0}(\infty) \bigg\rangle = \kappa c^2_{12} \delta^{ab} \not= 0.
}
While that of ${\cal A}_{oooc}$ also vanishes on-shell. So, $S$-matrix is almost trivial and 
therefore the theory is almost topological. 

\subsec{Target Space Low Energy Effective Action}

Non-vanishing three-point function on the sphere determines the low energy
effective action in the three cases previously considered.

For closed strings we have that Eq. \threep\ gives rise to an effective
field theory for a complex scalar field $\Omega$ whose dynamics is given
by the Pleba\'nski action \hh\

\eqn\plebact{
S_P = \int_{M^{2,2}} \bigg( \partial \Omega \bar{\partial} \Omega + {1 \over 3} \partial
\bar{\partial} \Omega \wedge \partial \bar{\partial} \Omega \bigg).
}
Equation of motion for this action is precisely the first heavenly
equation written in a slightly different form

\eqn\eompleban{
\partial^i \bar{\partial}_{\bar{i}} \Omega - 2 \kappa \partial \bar{\partial} \Omega \wedge \partial 
\bar{\partial} \Omega = 0.
}
Here, the complex scalar field $\Omega$ can be regarded as a perturbation
of the K\"ahler potential around the flat space $M^{2,2}$. The metric of
$M^{2,2}$ is given by $g_{i \bar{j}} = \partial_i \bar{\partial}_{\bar{j}}
(x_k \bar{x}^{\bar{k}} + 4 \kappa \Omega)$. This metric satisfies the
condition $\det (g_{i \bar{j}}) = -1.$ Therefore the corresponding Ricci
tensor is given by $R_{i \bar{j}} = \partial_i \bar{\partial}_{\bar{j}} (
log |\det g_{i \bar{j}}|) =0$. Thus the first heavenly equation is
equivalent to the Ricci-flatness condition. By Atiyah-Hitchin-Singer
theorem \ahs, a four-dimensional Riemannian space is self-dual if and
only if it is K\"ahler and Ricci-flat. Then, the space $M^{2,2}$ should be
a K\"ahler and Ricci-flat complex manifold.

For the open string case, the low energy effective action is given by

\eqn\openact{
S^o_{eff} = \widetilde{S}^o_{eff} + \int d^4x \bigg( -{g^2 \over 6} f^{adx}f^{xbc} \partial^i 
\varphi^a \cdot \varphi^b \cdot \overline{\partial}_{\bar{i}} \varphi^c \cdot \phi^d \bigg),
}
where

\eqn\effact{
\widetilde{S}^o_{eff} = \int d^4x \bigg( {1\over 2} \partial^i \varphi^a \cdot 
\overline{\partial}_{\bar{i}} \varphi^a - i {g \over 3} f^{abc} \varphi^a \cdot \partial^i 
\varphi^b \cdot \overline{\partial}_{\bar{i}} \varphi^c \bigg) + {\cal O}(\varphi^4).
}
The equation of motion of this action is precisely the Yang equation
\yang\

\eqn\yang{
\overline{\partial}_{\bar{i}} \bigg( \exp(-2ig \varphi) \partial^i \exp(2ig \varphi)\bigg)
=0, }
where $\varphi$ is a hermitian matrix. Yang equation is equivalent to
self-dual Yang-Mills 
equation $\widetilde{F}_{ij} = + F_{ij}.$

Up to here we have selected the option $(i)$ which consist of two
${\cal N}=2$ chiral
supermultiplets which give rise to a self-dual geometry
of the target space. In the case we select the option $(ii)$, that is, one chiral
superfield $X$ 
and one twisted chiral superfield $\Sigma$, we get a non-K\"ahlerian target space and
therefore a hermitian
self-dual manifold with torsion \refs{\gatestwo,\hull}. The torsion is generated by the
anti-symmetric
$B$-field. The case when this field vanishes leads to a free torsion hermitian
self-dual
manifold. In
this case the perturbative analysis of the scattering amplitudes for the action $S = \int 
{d^2 z\over \pi} d^4 K_0(X, \Sigma)$ might lead us to the {\it new} version of the heavenly
equation found 
recently by Pleba\'nski and Przanowski \new\

\eqn\newh{
\Omega_{,pr} \Omega_{,qs} - \Omega_{,ps} \Omega_{,qr} - {1 \over (1 - iqs)^2} \big( 2 \Omega 
\Omega_{,pr} -
\Omega_{,p} \Omega_{,r}\big) = 0,
}
where $\Omega = \Omega (p,q,r,s)$ with $x^1=p$, $x^2=q$, $\sigma^1=r$ and $\sigma^2=s.$

\noindent
[Remark\foot{I would like to thank Yaron Oz for pointing out to me these results.}: The
computation
of the effective
action arising from a world-sheet action, given
by ${\cal N}=2$ sigma model with chiral and twisted chiral superfields, was done by
Gluck, Oz and Sakai in Ref. \aefectiva. The result is the Laplace equation.]

\newsec{Hierarchy of Conserved Quantities}

All properties of the target space have a worldsheet interpretation. For instance, the
infinite hierarchy of conserved quantities in self-dual gravity are interpreted in terms of
the spectral flow of the a ground ring of ghost number zero operators in the chiral BRST
cohomology of the closed ${\cal N}=2$ string theory \giveon.

\subsec{Picture Changing, Spectral Flow and Ground Ring Structure}

It was found in Ref. \stringh\ a deep interrelation between the hidden
symmetries of self-dual gravity generated by abelian symmetries and the
{\it ground ring} of ghost number zero operators coming from the
BRST-cohomology. These operators are, in fact, related to the picture
changing operators and to the spectral flow.  In this subsection we
briefly review that construction (for details see Ref. \stringh).

Within the Superconformal Algebra (SCA) structure one can construct a BRST
current $j_{BRST}$ for the left-moving SCA and $\overline{j}_{BRST}$ for
the right-moving one. These currents are written in terms of the
generators $(T,J,G)_L$ and $(\overline{T},\overline{J},\overline{G})_R$ of
the corresponding SCA. Thus, one can define the BRST charge $Q_{BRST}$ as
$ Q_{BRST} = \oint \big( j_{BRST} + \overline{j}_{BRST}\big)$. Each ${\cal
N}=2$ SCA has two sets of picture families of charges $(\pi_+,\pi_-)$ and
$(\overline{\pi}_+,\overline{\pi}_-)$ connecting the families of Fock
spaces and giving rise to different {\it pictures}. The first systematic study 
of the BRST cohomology of ${\cal N}=2$ strings with picture changing was done in Ref. 
\bischoff. Another necessary
ingredient is the spectral flow operator $S$ which interpolate between the
$NR$ and $R$ sectors of the theory acting independently on left- and
right-movers. The ${\cal N}=2$ spectral flow operator was first constructed 
explicitly in Ref. \nsfketov.

Closed ${\cal N}=2$ strings needs from subsidiary constraints $(b_0 -
\bar{b}_0) |\psi \rangle =0$
and $b_0'|\psi \rangle = 0 = \bar{b}_0'|\psi \rangle$ with $(\pi_+, \pi_-)
\in \IZ \times
\IZ$.  Then, physical states are organized into cohomology classes called {\it relative
chiral} BRST-cohomology $H^*_{BRST}(g,\pi)$, where $g$ is the ghost number and $\pi$ is the
picture changing charge. The special case when the ghost number is zero $H^*_{BRST}(g=0,\pi)$
has the structure of a {\it ring} and it is called the {\it ground ring} \gr. This ring
structure is due to the fact that the multiplication of cohomology classes preserves the
ghost number and the picture gradings.  This product of cohomology classes is translated into
the OPE of operators. In this product a typical element of the the ground ring is given by

\eqn\ope{
{\cal O}^{\ell}_{m,n} = (X_+)^{j+m} \cdot (Y_-)^{j-m} \cdot H^{\ell},
}
where $\ell =0,1, \dots , 2j$, $m, n-m = -j, j+1, \dots , =j$ and $j=0,{1\over
2},1,{3\over2}, \dots$.  Here $X_+$, $X_-$ and $Y_-$ are suitable picture-raising operators
satisfying $X_+ \cdot H = X_- \cdot S$ and $Y_- = X_+ \cdot S^{-1} = X_- \cdot H^{-1}$ with
$H$ being a picture-neutral formal operator and $S$ being the spectral flow operator. The
changing picture operators $\Pi_{\pm}$ are playing the role of derivations in the polynomial
algebra of two-variable functions $x,y$. Thus defining

\eqn\coordxy{
x \equiv X_+ = {\cal O}^0_{{1\over 2}, +{1\over 2}}, \ \ \ \ \ \ \ \ 
y \equiv X_- = Y_- \cdot H = {\cal O}^0_{{1\over 2}, -{1\over 2}} \cdot H.
}
In terms of these variables we have

\eqn\operdos{
{\cal O}^{\ell}_{m,n} = x^{j+m} \cdot y^{j-m} \cdot h^{-(j-m) + \ell}, \ \ \ \ \ 
\Pi_+ +1 = x {\partial \over \partial x},  \ \ \ \   \Pi_- +1 = y {\partial \over \partial
y}.
}

\vskip .5truecm
\noindent
{\it Geometric Structure of the Ground Ring Manifolds}

As we have seen before the physical states of the theory generate the
ground ring structure which encodes all relevant information about the
symmetries of the theory. Moreover, much of the power lies in that
actually there is a geometrical structure underlying the BRST approach.
This is precisely the symplectic geometry \wz\ and the theory of homotopy
Lie algebras \lz. Let $C^{\infty}({\cal A}_L)$ be the chiral ground ring
with the operators ${\cal O}_{u,n} = h \cdot x^{u+n} \cdot y^{u-n}$ are
precisely the polynomial functions on the $x-y$-plane ${\cal A}_L$. This
is a two-dimensional symplectic manifold with symplectic two-form $\omega
= dx \wedge dy.$ Thus, the pair $({\cal A}_L, \omega)$ is a
two-dimensional symplectic manifold with the symplectic two-form $\omega$.
Similarly we can say the same about the right-movers, i.e., $({\cal
A}_R,\widetilde{\omega})$, where $\widetilde{\omega} = d \bar x \wedge d
\bar y$.

Now we review briefly the result of \groundring. Given ${\cal A}_L$ and
${\cal A}_R$ we define the {\it full quantum ground ring manifold} ${\cal
W}$ to be the product $ {\cal W} = {\cal A}_L \times {\cal A}_R.$ Define
$\rho: {\cal A}_L \times {\cal A}_R \to {\cal A}_L$ and $\widetilde{\rho}:
{\cal A}_L \times {\cal A}_R \to {\cal A}_R$ to be the corresponding
projections. Operators $\Pi_{\pm}$ can be interpreted as Hamiltonian (or
volume-preserving) vector fields on the the full quantum ground ring
manifold ${\cal W}$.

\vskip 1truecm
\subsec{Boyer-Pleba\'nski Construction}

In this subsection, we re-derive the same structure in a different way
making emphasis only on the fact that the full quantum ground ring
manifold ${\cal W}$ is merely the product manifold ${\cal A}_L \times {\cal
A}_R$, and
that the chiral ground ring manifolds are symplectic manifolds with symplectic
two-forms $\omega = dx \wedge dy $ and $\widetilde{\omega} = d\bar x \wedge d\bar
y$ respectively. For this end we use the construction of \bpone\ and \bptwo.

\vskip 1truecm
\noindent
{\it The Construction}

Consider the flat chiral complexified ground ring manifolds ${\cal
A}_L^{\IC}$ and   ${\cal A}_R^{{\IC}}$ and the complexified full
ground ring manifold $ {\cal W}^{{\IC}} = {\cal A}_L^{{\IC}}\times
{\cal A}_R^{{\IC}}$. Since $({\cal A}_L^{{\IC}}, \omega)$ and
$({\cal
A}_R^{{\IC}}, \widetilde{\omega})$ are symplectic manifolds one can
show that  ${\cal W}^{{\IC}}$ is also a symplectic manifold with symplectic
form given by $\rho^* \omega - \widetilde{\rho}^* \widetilde{\omega}$  where $\rho\ : {\cal
W}^{{\IC}} \to {\cal A}_L^{{\IC}} $ and $\widetilde{\rho} \ : {\cal
W}^{{\IC}} \to {\cal A}_R^{{\IC}} $ are the corresponding projections.

Now, let  $T^r{\cal A}^{{\IC}}_L$, $T^r{\cal A}^{{\IC}}_R$ and $T^r{\cal
W}^{{\IC}}$ be the $r$-th order holomorphic tangent bundles  of ${\cal
A}^{{\IC}}_L$, ${\cal A}^{{\IC}}_R$ and ${\cal W}^{{\IC}}$,
respectively. Then we have the following sequence of projections for
${\cal A}^{\IC}_L$ as follows:

\eqn\sucesion{
 \dots \to T^r{\cal A}^{\IC}_L \to T^{r-1}{\cal A}^{\IC}_L \to
\dots \to T^1{\cal A}^{\IC}_L \to T^0{\cal A}^{\IC}_L \cong {\cal A}^{\IC}_L}
and similarly for ${\cal A}^{\IC}_R$ and ${\cal W}^{\IC}$.

Following \bptwo\ one can define functions ${\cal O}^{(\lambda)}_{u,n}$,
$\bar {\cal O}^{(\lambda ')}_{u,n'}$ and ${\cal V}_{u,n,n'}^{(\lambda)}$;  
vector fields $Y^{+(\lambda)}_{s,n}$, $\bar Y^{+(\lambda ')}_{s,n}$ as
well as ${\cal J}_{u,n,n'}^{(\lambda)}$, $\bar{\cal J}_{u,n,n'}^{(\lambda
')}$, and differential forms $\omega
^{(\lambda)}$, $\widetilde{\omega}^{(\lambda)}$, (with $\lambda =
0,1,2,\dots,r$) on the $r$-order holomorphic tangent (or cotangent)
bundles $T^r{\cal A}^{\IC}_L$, $T^r{\cal A}^{\IC}_R$ and $T^r{\cal
W}^{\IC}$. For example ${\cal O}^{(\lambda)}_{u,n}(j_r\circ \psi(0)) =
{1\over \lambda !}{d^{\lambda}({\cal O}_{u,n}\circ \psi) \over
dt^{\lambda}} \mid _{t=0}$ and $ {\cal V}^{(\lambda)}_{u,n,n'}(j_r\circ
\psi(0)) = {1\over \lambda !}{d^{\lambda}({\cal V}_{u,n,n'}\circ \psi)
\over dt^{\lambda}}\mid_{t=0} $, where $j_r(\psi)$, is the $r$-jet of the
holomorphic curve $\psi$. Then we can define the OPE for these objects as

\eqn\operadores{
{\cal O}^{(\lambda)}_{u,n}(j_r\circ \psi (0))\cdot {\cal
O}^{(\lambda')}_{u,n'}(j_r\circ \psi (z)) = {1\over \lambda !}
{\partial^{\lambda}({\cal O}_{u,n}\circ \psi) \over \partial s^{\lambda}} \mid
_{s=0}\cdot {1\over \lambda ' !}{\partial^{\lambda'}({\cal O}_{u,n'}\circ
\psi) \over \partial t^{\lambda'}} \mid _{t=z}.
}

From the theorem 2 of \bptwo\ one can see that $(T^r{\cal A}^{\IC}_L,
\omega^{(\lambda)})$ and $(T^r{\cal A}^{\IC}_R,
\widetilde{\omega}^{(\lambda)})$ are symplectic manifolds. Therefore, we
can define another symplectic manifold $T^r{\cal W}^{\IC}$ with symplectic
two form given by $\rho^*\omega^{(r)} -
\widetilde{\rho}^*\widetilde{\omega}^{(r)}$. One can also establish the
bundle diffeomorphism ${\cal Q}: T^r{\cal W}^{\IC} \to \rho^*T^r{\cal
A}^{\IC}_L \oplus \widetilde{\rho}^* T^r{\cal A}^{\IC}_R.$

The following diagram summarizes our construction:

$$
\matrix {T^2{\cal W}^{\IC}  &\to   &\rho^*T^2{\cal
A}^{\IC}_L\oplus\widetilde{\rho}^*T^2{\cal A}^{\IC}_R &
\cr
\pi^2_1 \downarrow      &      & \downarrow \rho_1^2 & \searrow \widetilde{\rho}^2_1&
\cr
T{\cal W}^{\IC} &&\rho^*T^2{\cal A}^{\IC}_L   \to   &\widetilde{\rho}
T^2{\cal A}^{\IC}_R
\cr
\pi \downarrow && &&&\cr
{\cal W}^{\IC} &&&& \cr} 
$$

In order to show the existence of self-dual gravity structures on the full
quantum ground ring manifold we restrict ourselves to the case $r=2$. As
it is mentioned in \bptwo\ $(\rho^* T^2 {\cal A}^{\IC}_L,
\rho^*\omega^{(2)} - \widetilde{\rho}^*\widetilde{\omega}^{(0)})$ and
$(\rho^*T^2{\cal A}^{\IC}_R, \rho^*\omega^{(0)} -
\widetilde{\rho}^*\widetilde{\omega}^{(2)})$ are also symplectic
manifolds, where $\omega^{(0)} = \omega = dx\wedge dy,$
$\widetilde{\omega}^{(0)} = \widetilde{\omega} = d\bar x\wedge d\bar y,$
$\omega^{(2)} = dx\wedge dy^{(2)} + dx^{(1)}\wedge dy^{(1)} +
dx^{(2)}\wedge dx,$ and $\widetilde{\omega}^{(2)} = d\bar x\wedge d\bar
y^{(2)} + d\bar x^{(1)}\wedge d\bar y^{(1)} + d\bar x^{(2)}\wedge d\bar
x.$ We need only to consider the manifold $\rho^*T^2{\cal A}^{\IC}_L$. 

Consider the following sequences

\eqn\succuno{
{\cal L}\buildrel{\rm i} \over{\to} T^2 {\cal W}^{\IC}\buildrel {\pi^2_1}
\over{\to} T{\cal W}^{\IC}\buildrel{\pi^1} \over{\to} {\cal W}^{\IC},
}
and
\eqn\succdos{ 
{\cal L}\buildrel\rm i \over{\to} T^2{\cal W}^{\IC}\buildrel\rm \rho^2_1
\over{\to} \rho^*T^2{\cal A}^{\IC}_L\buildrel\rm \rho^1 \over{\to} {\cal W}^{\IC},
}
where ${\cal L}$ is a horizontal Lagrangian submanifold of both $T{\cal W}^{\IC}$ 
and $\rho^*T^2{\cal A}^{\IC}_L$.

Let $\sigma : W \to T^2{\cal W}^{\IC},\ W \subset {\cal W}^{\IC}$, be a holomorphic
section such that $i({\cal L}) = \sigma(W)$, where $i: {\cal L} \to T^2{\cal
W}^{\IC}$
is an injection.

Let $\Gamma(W)$ be the set of all holomorphic sections of $T^2{\cal
W}^{\IC} \to {\cal W}^{\IC}$ such that $\pi_1^2\circ i({\cal L})$ and
$\rho^2\circ i({\cal L})$ are horizontal Lagrangian submanifolds of
$T{\cal W}^{\IC}$ and $\rho^*T^2{\cal A}^{\IC}_L$, respectively. Thus
$\Gamma(W)$ is defined by the equations

\eqn\seccionuno{
\sigma_1^*(\omega^{(1)} - \widetilde{\omega}^{(1)}) = 0, \ \ \ {\rm on} \
\ W \subset {\cal W}^{\IC},
}

\eqn\secciondos{
\sigma_2^*(\omega^{(2)} - \widetilde{\omega}^{(0)}) = 0, \ \ \ {\rm on} \
\ W\subset {\cal W}^{\IC},
}

\noindent
[Notice that we have omitted the pull-back $\rho^*$ and $\widetilde{\rho}^*$
in these formulas.]

\noindent
where $\omega^{(1)} = dx \wedge dy^{(1)} + dx^{(1)}\wedge dy$,  $\widetilde{\omega}^{(1)} 
= d\bar x \wedge d\bar y^{(1)} + d\bar x^{(1)}\wedge d\bar y$,
$\sigma_1 = \pi_1^2 \circ \sigma$ and $\sigma_2 = \rho^2\circ \sigma$.

Now the problem arises of {\it how $\sigma \in \Gamma(W)$
determines  a
self-dual gravity structure on the open sets $W$ of the full quantum
ground ring manifold ${\cal W}^{\IC}$. }

The solution of this problem is given by the following

\noindent
{\it Theorem \bptwo:} Let  $\sigma \ : \ W \to T^2{\cal W}^{\IC}, \
W \subset {\cal W}^{\IC}$, be a holomorphic section. The triplet
$(\omega, \widetilde{\omega}, \Omega_0) =(\sigma^* \omega^{(0)}, \sigma^*\widetilde
{\omega}^{(0)}, \sigma^* \omega^{(1)})$ defines a self-dual structure
on $W$ if and only if there exist a choice of a holomorphic section
$\sigma$ such that $\sigma^*(\omega^{(1)} - \widetilde{\omega}^{(1)}) =0$ and
$\sigma^*(\omega^{(2)} - \widetilde{\omega}^{(0)}) = 0$. 

\noindent
(For the proof see \bptwo).
Thus, the desired self-dual gravity structures arise in a natural
manner from the mathematical structure of the quantum states in ${\cal
N}=2$ string theory.

Taking $\Omega_0 = \sigma^*_1 [dx\wedge dy^{(1)} + dx^{(1)} \wedge dy]$ we
arrive at the first heavenly equation

\eqn\heavens{
\Omega_0\wedge\Omega_0 +  2\omega\wedge \widetilde{\omega} = 0. 
}
This can be extended to the cases with $r\geq 3 $. Then, by using the
projective limit one can formulate the problem in terms of the
infinite-dimensional tangent bundle $T^{\infty} {\cal W}^{\IC} =
\rho^*T^{\infty}{\cal A}^{\IC}_L \oplus \widetilde{\rho}^* T^{\infty}{\cal
A}^{\IC}_R$ (for details see \bptwo). It can be proved that given $({\cal
W}^{\IC}, \rho^*\omega - \widetilde{\rho}^* \widetilde{\omega})$ a
symplectic manifold $(T^{\infty}{\cal A}^{\IC}_L, \omega_2(t))$ turns out
to be a formal symplectic manifold, where $\omega_2 (t) =
\sum_{k=0}^{\infty} \pi_k^*\omega^{(k)}t^k$;  $t\in {\IC}$, and $ \pi_k \
: \ T^{\infty}{\cal A}^{\IC}_L \to T^k{\cal A}^{\IC}_L$ is the natural
projection.

By the Proposition 2 of \bptwo\  we observe that  $(T^{\infty} {\cal
W}^{\IC},
\omega(t))$ is a formal symplectic manifold with

\eqn\combina{
\omega(t)= t^{-1} \rho^* \omega_2(t) - t \widetilde{\omega}_2(t^{-
1}),
}
where $t\in {\IC}^*\equiv {\IC} - \{0\}$ and $T^{\infty}{\cal W}^{\IC} =
T^{\infty}{\cal A}^{\IC}_L \times T^{\infty} {\cal A}^{\IC}_R =
T^{\infty}({\cal A}^{\IC}_L \times {\cal A}^{\IC}_R) = \rho^*T^{\infty}
{\cal A}^{\IC}_L \times \widetilde{\rho}^* T^{\infty}{\cal A}^{\IC}_R.$

\vskip .5truecm
\noindent
{\it Curved Twistor Construction on Full Quantum Ground Ring Manifolds}

Consider the formal symplectic manifold $(T^{\infty}{\cal W}^{\IC},
\omega(t))$. Since  ${\cal A}^{\IC}_L$ and ${\cal A}^{\IC}_R$ are
diffeomorphic, we have $T^{\infty}{\cal W}^{\IC}= T^{\infty}{\cal
A}^{\IC}_L\times T^{\infty}{\cal A}^{\IC}_R$. Define the holomorphic maps

\eqn\disco{
\widehat{\cal D}= (D,I) : T^{\infty}{\cal A}^{\IC}_L \times {\IC}^* \to
T^{\infty}{\cal A}^{\IC}_L \times {\IC}^*, 
}
where $I(t) =t^{-1}$ and the graph of the diffeomorphism $D$, $grD$, can be
identified with some local section $grD=\sigma ' : W \to T^{\infty}{\cal
W}^{\IC}$ such that $\sigma '^* \omega(t) = 0$. From Eq. \combina,
this last
relation holds if and only if

\eqn\discodos{
D^*\omega_2(t^{-1}) = t^{-2} \omega_2(t).
}

Consider now a local section  $\sigma ''$ of the formal tangent bundle
$T^{\infty} {\cal W}^{\IC} \to {\cal W}^{\IC}$ on a open set $W \subset
{\cal W}^{\IC}$ such that $\sigma ''^* \omega(t) =0$. For $t\in
{\IC}^*$ we have $\sigma ''= (\Psi^A(t),\widetilde{\Psi}^B(t^{-1})).$
Now assume that $\Psi^A(t)$ and $\widetilde{\Psi}^B(t^{-1})$ converge in
some open
disks ${\cal U}_0$ and ${\cal U}_{\infty}$ ($0\in {\cal U}_0$ and $\infty \in
{\cal U}_{\infty})$ respectively, such that ${\cal U}_0 \cap {\cal
U}_{\infty} \not= \phi$. Consequently, the functions $\Psi^A: t\mapsto
\Psi^A(t)$ and $\widetilde{\Psi}^B: s\mapsto \widetilde{\Psi}^B(s)$ define local
holomorphic sections of the twistor space ${\cal T}$. Due to the condition
\discodos\ defining the self-dual structure on the quantum ground ring
manifold
${\cal W}^{\IC}$ we get the transition functions for a global holomorphic
section $\Psi \in \widetilde{\Gamma}({\cal T}).$ Thus one can recover the
Penrose twistor construction \penrose. Of course the inverse process is also possible.

\newsec{Mirror Symmetry}

The theory of ${\cal N}=2$ strings also presents many features of
Calabi-Yau three-folds $X$. One of them is {\it mirror symmetry}. However
this symmetry is realized in a different form than in Calabi-Yau
three-folds $X$. For CY-3 fold mirror pair $(X,Y)$ the fact that the Betti
number $h^{2,0}(X)=0$ leads to a local factorization of the moduli space
of K\"ahler ${\cal M}_{h^{1,1}(X)}$ and complex structures ${\cal
M}_{h^{2,1}(Y)}$ i.e., ${\cal M} = {\cal M}_{h^{1,1}(X)} \times {\cal
M}_{h^{2,1}(Y)}$.  Moreover, in topological sigma models on a CY 3-fold
there are two possible ways to twist these class of models. These are the
${\cal A}(X)$ and ${\cal B}(X)$ models \mirror. Mirror symmetry is
realized through the interchanging of ${\cal A}(X)$ and ${\cal B}(X)$
models.

In (0,4)  real four-dimensional mirror pairs $(\widetilde{X},
\widetilde{Y})$, for instance, for the $\widetilde{X}=K3$ surface this is
not true, since $h^{2,0}(\widetilde{X})=1$, the moduli spaces are mixed
and it cannot be factorized \aspin. However the mirror map still
interchanges the moduli of K\"ahler and complex structures. At the level
of the metric, two metrics $g(\widetilde{X})$ and $g^*(\widetilde{Y})$,
under the mirror map, are related by $g^* = \phi^*(g)$, with $\phi$ a
non-trivial automorphism of the moduli space of metrics. For the present
case, i.e., in the non-compact case with signature (2,2), this
automorphism corresponds to a change of gauge which gives the two
different metrics which are solutions of the first and the second heavenly
equations \fhe\ and \she. Thus first and second heavenly equations are
related by a mirror map of a non-compact CY two-fold or self-dual space
i.e., $(\widetilde{X}, g_{\Omega})  {\to} (\widetilde{Y}, g_{\Theta})$.
Thus, we have

\eqn\mmap{
{\cal A}(\widetilde{X}, g_{\Omega}) \to {\cal B}(\widetilde{Y},
g_{\Theta}),
}
where $\Omega$ and $\Theta$ are solutions of the first and the second heavenly equations
\fhe\ and \she\ respectively.  These considerations are confirmed both for
closed and open
strings by the explicit computation of the worldsheet instantons in
Ref. \inst.

For open strings in the ${\cal A}$ model it gives the Leznov-Parkes
equations \lezparkes\ (instead
of the Yang's equation \yang) and for the closed string it gives rise to
the second heavenly equation \she. In what follows we will consider only
the ${\cal A}$-twist and we will give some explicit examples of the so
called {\it large}-$N$ {\it duality} which is a manifestation of the {\it
open/closed
duality} \refs{\cswitt,\gopa} (see \mmarino, for a recent review). Thus,
we only
will consider issues related to the second
heavenly equation \she.

\ifig\mirror{The figure represents the mirror equivalence between the type ${\cal A}$ and
${\cal B}$ topological ${\cal N}=2$ strings models. This gives rise to a
stringy
explanation of the
equivalence of
the first and second heavenly equations.} 
{\epsfxsize4in\epsfbox{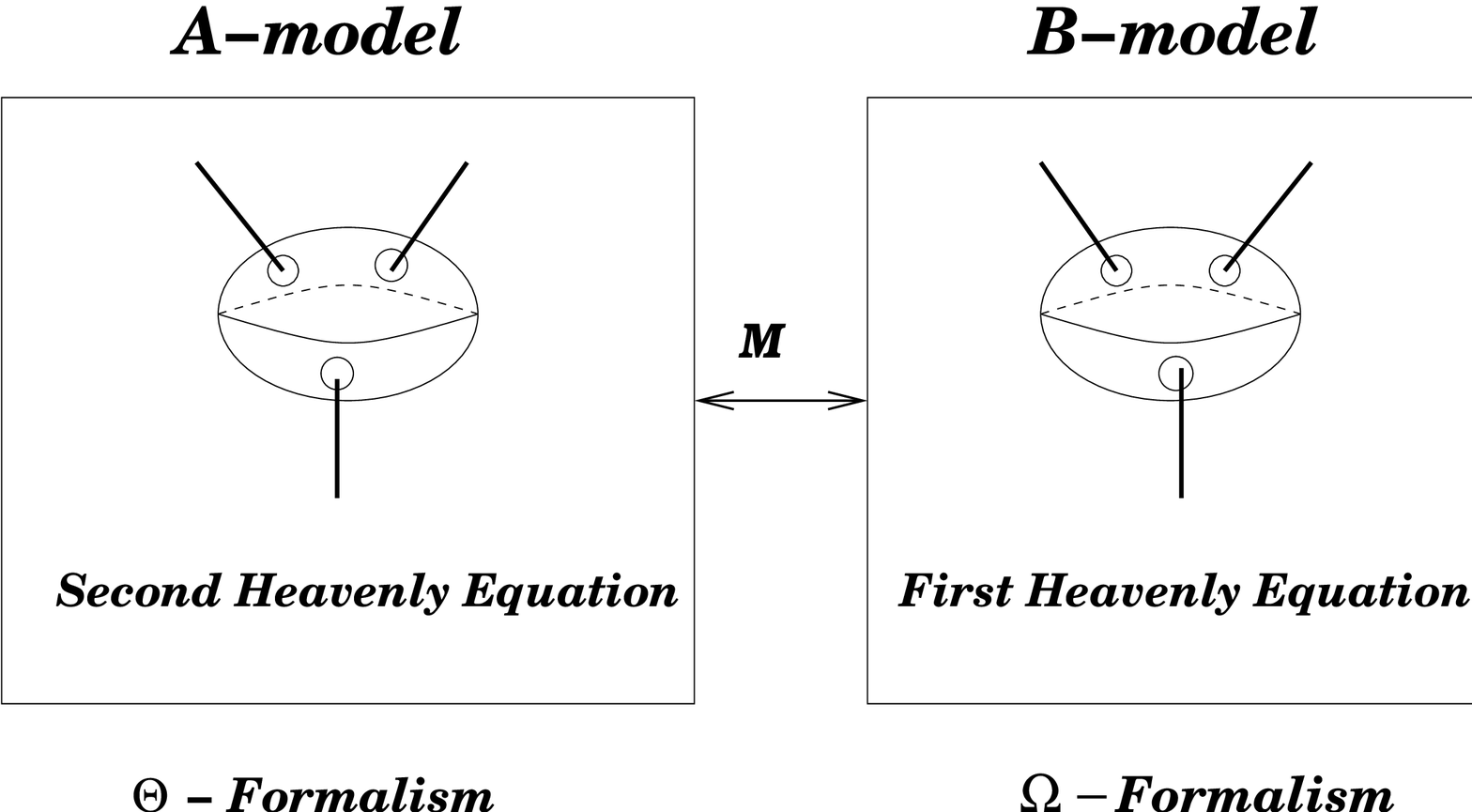}}


\newsec{Large $N$ Duality  and Open/Closed Duality}

For CY 3-folds also there are nice correspondences besides of the mirror
symmetry. For ${\cal A}$ and ${\cal B}$ models there is an internal
correspondence between open and closed topological strings. This is a
topological open/closed duality which is also known as {\it large
N-duality}.  For instance, in ${\cal A}$ models it interchanges (from the
open string sector) the {\it Chern-Simons gauge theory} with the {\it K\"ahler
gravity} (in the closed string sector) \kg. For ${\cal B}$ models it interchanges
{\it holomorphic Chern-Simons gauge theory} from the open sector with {\it
Kodaira-Spencer gravity} from the closed one. This duality has refinements
once that one incorporates topological D-branes. Gopakumar and Vafa \gopa\
showed that the open sector can be regarded as topological D-branes
wrapping a {\it deformed conifold} and this sector is {\it dual} to a
topological closed string (without D-branes) propagating on the {\it
resolved conifold}.

For non-compact CY 2-folds this duality is also true. One example
is precisely, the ${\cal N}=4$ topological open string. We start with the
Park-Husain heavenly equation for the holomorphic function $\Theta
=\Theta(x,y,p,q)$, describing self-dual spacetime \chiral\

\eqn\ph{
\partial^2_x \Theta + \partial^2_y \Theta  + \{\partial_x
\Theta,\partial_y \Theta\}_P =0,
}
where the Poisson bracket is given by $\{F,G\}_P =\big({\partial F\over
\partial p} \cdot {\partial G\over \partial q} - {\partial F\over \partial
q} \cdot {\partial G\over \partial p} \big)$. On the self-dual space
$M^{2,2} = {\cal X} \times {\cal Y}$ with $(x,y)$ and $(p,q)$ the local
coordinates on ${\cal X}$ and ${\cal Y}$ respectively and $\Theta(x,y,p,q)
= \Theta(x^i,y^i)$. Equation \ph\ can be derived from the Lagrangian
$S_{\infty} = \int_{{\cal X} \times {\cal Y}} dx dy dp dq {\cal
L}_{\infty}$ where the Lagrangian is given by

\eqn\lagph{
{\cal L}_{\infty} = {1\over 2} \bigg( (\partial_x \Theta)^2  + (\partial_y \Theta)^2\bigg)  - {1 
\over 3} \Theta  \{\partial_x \Theta, \partial_y \Theta \}_P.
}

Equation of motion \ph\ are classically equivalent\foot{These systems are classically equivalent 
but at the quantum level they have very different properties \nappi.}
 to the following system

$$
F_{xy} = \partial_x A_y - \partial_y A_x + \{A_x,A_y\}_P = 0,
$$

\eqn\flatconn{
\partial_xA_x + \partial_y A_y = 0, \ \ \ \ \ \ \ \ A_x = - \partial_y \Theta, \ \ \ \  A_y =
\partial_x \Theta.
}
We can identify a sdiff$({\cal Y})$-valued {\it flat} connection. As a Lie algebra
sdiff$({\cal Y})$ is isomorphic to su$(\infty)$. Thus we have a large-$N$ limit of a
two-dimensional gauge theory corresponding to the flat connection. At the classical level we
may have equivalently the system

$$
\partial_x(g^{-1} \partial_x g) + \partial_y(g^{-1} \partial_y g) = 0,
$$

\eqn\chiral{
F_{xy} = 0, \ \ \ \ \ \ A_x = g^{-1} \partial_x g, \ \ \ \  A_y = g^{-1} \partial_y g.}

Equations \chiral\ can be obtained from  the Lagrangian

\eqn\lagchiral{
I = \int_{\cal X} {\rm Tr}(g^{-1}dg)^2,}
where ${\rm Tr}(\cdot) = - \int_{\cal Y} dp dq (\cdot).$

For finite group, this result can be interpreted as the effective field theory on a
topological D1 brane wrapped in the two-sphere ${\cal X}={\bf S}^2$ in the target space
${\cal X} \times {\cal Y} =
T^* {\bf S}^2$. This is precisely the deformed conifold.

Something similar can be carried over to the six-dimensional version of
the heavenly equation \plebprzan\ for the holomorphic function
$\Theta(x,y,\widetilde{x},\widetilde{y},p,q)$

\eqn\six{
\partial_x \partial_{\widetilde{x}} \Theta + \partial_y \partial_{\widetilde{y}} \Theta
+ \{\partial_x \Theta, \partial_y \Theta\}_P = 0.}

\newsec{D-branes in ${\cal N}=2$ Strings}

In this section we briefly survey the problem of having D-branes in ${\cal N}=2$
strings. This is a very interesting and important topic nowadays.

It is known that ${\cal N}=2$ strings has not space-time supersymmetry.
Thus, the usual definition of D-branes as BPS states is not longer valid.
However we still have the notion of space-time self-duality which, in many
respects, plays the role of a BPS condition.

In Ref. \martinec, it was argued that the effective field theory on a
``D-brane'' of ${\cal N}=2$ string theory is determined by dimensional
reductions
of the self-duality Yang-Mills theory on $M^{2,2}$. The different reduced
theories give rise to an spectrum of integrable systems in various
dimensions. For instance in dimension three it is given by the Bogomolny
equation. In two dimensions it is given by the Hitchin system, in one
dimension by the Nahm equation and in zero dimensions by the ADHM
equation, In the literature it was discovered that distinct ${\cal N}=2$ must arise from
the  existence of models that use mixtures of chiral and twisted chiral multiplets
\gatesthree.

Very recently, in Refs. \refs{\cheung,\aefectiva}, this fact was rediscovered in the
context of the conformal field theory formalism. Thus ${\cal N}=2$ theories were
classified by families of theories according whether the left and right
SCFA has different (inequivalent) complex structures. This give
rise to $\alpha$-strings and $\beta$-strings respectively. It was proved
there that both families are related by T-duality. 

In Ref. \refs{\aefectiva,\dbranes}, the
D-brane-D-brane
scattering amplitudes were studied in more detail, as well as D-brane scattering with
open and closed
strings in different cases and then the brane effective field
theory on the brane of different dimensions was computed giving rise precisely to the
field theories argued in \martinec. Also in \cheung\ the coupling of
D-branes to the bulk theory for $\alpha$ and $\beta$-strings was computed.

\ifig
\closedopen{ The figure
represents the equivalence between the gauge theory on infinity number of topological
D-branes on the resolved conifold and a gravity theory (Park-Husain theory) on the deformed
conifold.}
{\epsfxsize3.8in\epsfbox{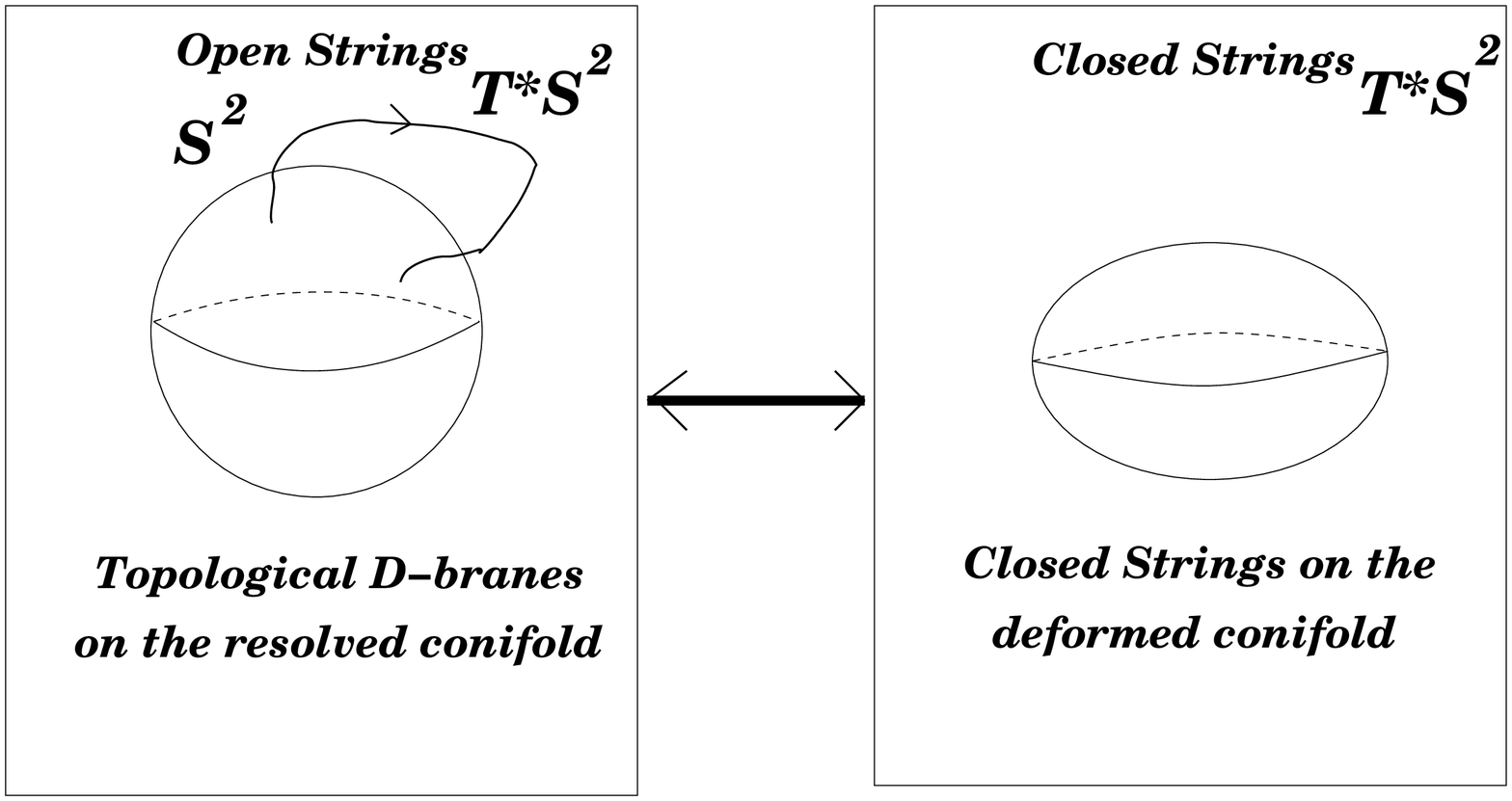}}


\newsec{Final Remarks}

In this contribution we have discussed the {\it target space/worldsheet}
correspondence of ${\cal N}=2$ strings. These theories are of topological
nature
and they are completely determined by its target space geometry, which is
a {\it self-dual} geometry. Thus self-dual geometry can be regarded as
an
avenue to study string theory and vice-verse. Thus self-dual geometry
would be
destinated to shed some light of what really string theory is. For this
reason, many results of those obtained by Pleba\'nski and his co-workers would be
important to continue exploring this {target space/worldsheet}
correspondence. One of these generalizations is the extension of self-dual
geometry to all algebraically degenerated ${\cal H}$-spaces and in
general, to know what is the stringy version of the Penrose-Petrov
classification. Also of particular interest is the stringy description of
${\cal HH}$-spaces. It is known that ${\cal HH}$-spaces under certain
conditions can give rise to {\it real} spacetimes \refs{\hhone,\hh} and it would be also a
bridge to connect ${\cal N}=2$ String theory to real space-time physics.
${\cal
HH}$-spaces are not self-dual spaces but they are so close to them that
the methods pursued by ${\cal N}=2$ string theory surely will be useful. Recently a new
${\cal N}=2$ supersymmetric extension of the heavenly equation was found in Ref.
\toppan. It would be very interesting whether this equation can be related to ${\cal
N}=2$ strings.

Finally it would be interesting to address, in the context of the present
paper, the correspondence between YM amplitudes and topological string in
twistor space according to Refs. \refs{\twistor,\berko}.


\vskip 30pt
\centerline{\bf Acknowledgments}

I would like to thank Prof. Jerzy Pleba\'nski to whom I am indebted for
many valuable suggestions and recommendations along my carrier. It is a
pleasure to thank him and Maciej Przanowski for a very enjoyable
collaboration for the years. Correspondence with S.J. Gates Jr., S. Ketov, J.P. Nunes,
Y. Oz and  F. Toppan is  warmly appreciated. This work has been supported by the grants
33951-E and 41993-F from Mexico's National Council of Science and
Technology (CONACyT). I want to thank also ICTP for hospitality during the
summer of 2003 in where this paper was written.

\listrefs
\end